\documentclass[10pt,b5paper,twoside]{ise2}
\usepackage{graphicx}
\usepackage{natbib}
\begin{document}
\title{Chemodynamical Modeling of Dwarf Galaxy Evolution}

\author{\inst{1}P. Berczik,~~\inst{2}G. Hensler,~~\inst{2}Ch. Theis and~~\inst{3}R. Spurzem \\
\institutename{\inst{1}Main Astronomical Observatory, Ukrainian National Academy of Sciences} \\
\instituteaddress{\inst{~}Zabolotnoho Str., 27, 03680, Kiev, Ukraine.}~\inst{~}\email{berczik@mao.kiev.ua} \\
\institutename{\inst{2}Institut f\"ur Theoretische Physik und Astrophysik, University of Kiel,} \\
\instituteaddress{\inst{~}Olshausenstr. 40, 24098 Kiel, Germany.} \\
\institutename{\inst{3}Astronomisches Rechen-Institut,} \\
\instituteaddress{\inst{~}M\"onchhofstra\ss e 12-14, 69120 Heidelberg, Germany.}}

\date{}

\maketitle

\begin{abstract}
\noindent We present our recently developed 3-dimensional
chemodynamical code for galaxy evolution. It follows the evolution of
all components of a galaxy such as dark matter, stars, molecular clouds
and diffuse interstellar matter (ISM). Dark matter and stars are
treated as collisionless $N$-body systems. The ISM is numerically
described by a smoothed particle hydrodynamics (SPH) approach for the
diffuse (hot) gas and a sticky particle scheme for the (cool) molecular
clouds. Additionally, the galactic components are coupled by several
phase transitions like star formation, stellar death or condensation
and evaporation processes within the ISM. As an example here we present 
the dynamical, chemical and photometric evolution of a star forming dwarf 
galaxy with a total baryonic mass of $2 \times 10^9 {\rm M}_\odot$.
\\
\keywords{computational methods: SPH, chemodynamics -- evolution of
galaxies: dwarf galaxy evolution}
\end{abstract}

\section{Introduction}

Since several years smoothed particle hydrodynamics (SPH, \cite{M1992})
calculations have been applied successfully to study the formation and
evolution of galaxies. Its Lagrangian nature as well as its easy
implementation together with standard $N$-body codes allows for a
simultaneous description of complex dark matter-gas-stellar systems
\citep{NW1993, MH1996}. Nevertheless, until now the present codes lack
of processes that are based on the coexistence of different phases of
the interstellar medium (ISM), mainly dissipative, dynamical and
stellar feedback, element distributions, etc. We have therefore
developed a 3d chemodynamical code which is based on our single phase
galactic evolutionary program \citep{Ber1999, Ber2000}.

\begin{figure}[t!]
\centerline{%
\begin{tabular}{c@{\hspace{0.1in}}c}
\includegraphics[width=2.35in]{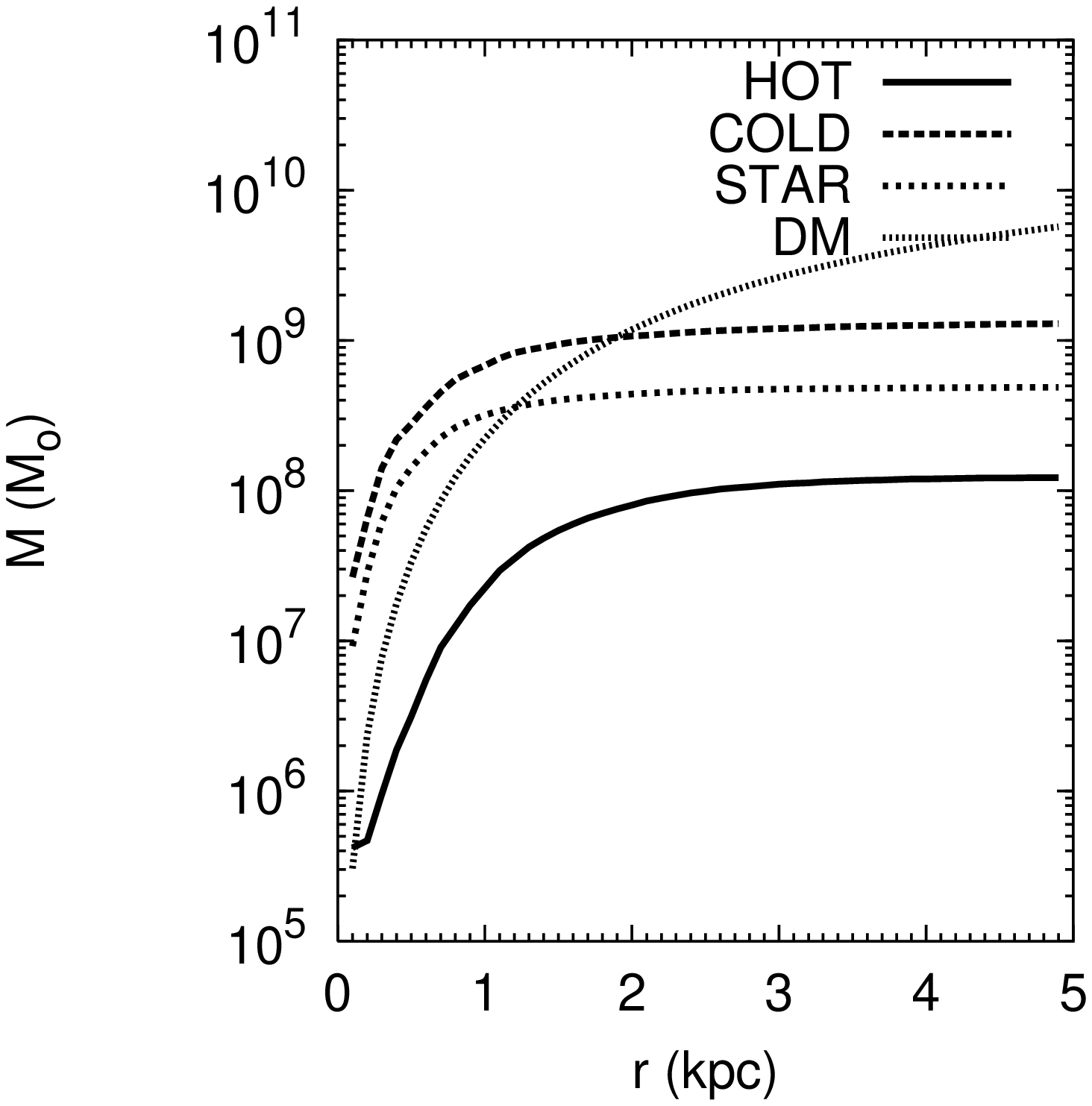} &
\includegraphics[width=2.35in]{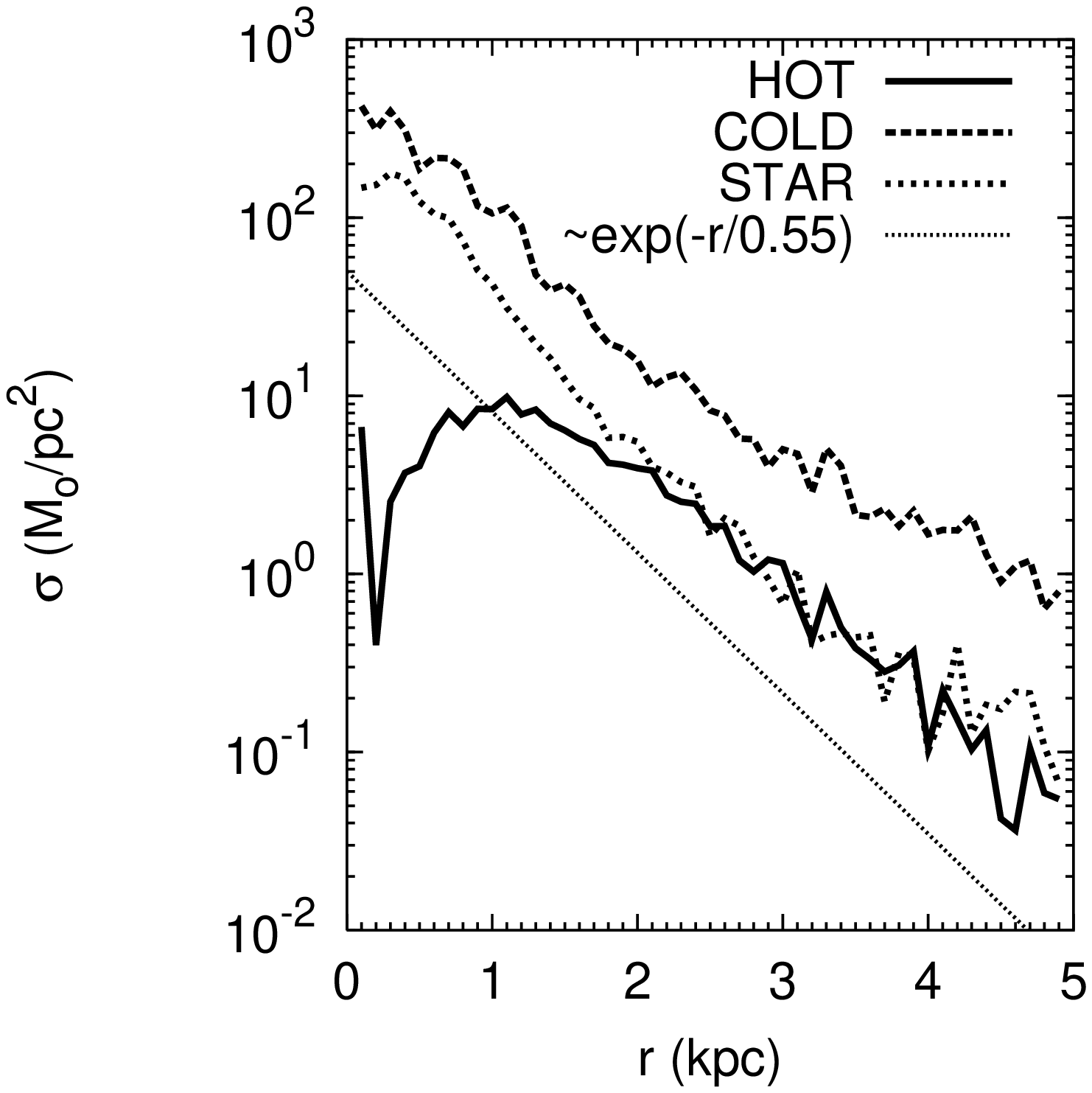}
\end{tabular}}
\caption{The radial distribution of the cumulative mass (left) and the
surface density (right) for the different components in the central
region of the model galaxy after 1~Gyr.}
\label{fig:100-mass&100-sigma}
\end{figure}

This code includes many complex effects such as a multi-phase ISM,
cloud-cloud collisions, a drag force between different ISM components,
condensation and evaporation of clouds (CE), star formation (SF) and a
stellar feedback (FB). The more detailed description of the new
code and the full list of the interaction processes between all gaseous
and stellar phases will be presented in a more comprehensive paper by
\cite{BHTS2002}. Here we just briefly describe some basic features and
effects.

In our new (multi-phase gas) code we use a two component gas
description of the ISM \citep{TBH1992, SHT1997}. The basic idea is to
add a cold (10$^2$ - 10$^4$ K) cloudy component to the smooth and hot
gas (10$^4$ - 10$^7$ K) described by SPH. The cold clumps are modeled
as $N$-body particles with some ``viscosity'' \citep{TH1993}
(cloud-cloud collisions and drag force between clouds and hot gas
component). The cloudy component interacts with the surrounding hot gas
also via condensation and evaporation processes \citep{CMcKO1981,
KTH1998}. In the code we introduce also star formation. The ``stellar''
particles are treated as a dynamically separate (collisionless)
$N$-body component. Only the cloud component forms the stars. During
their evolution, these stars return chemically enriched gas material
and energy to both gaseous phases.

\begin{figure}[t!]
\centerline{%
\begin{tabular}{c@{\hspace{0.1in}}c}
\includegraphics[width=2.35in]{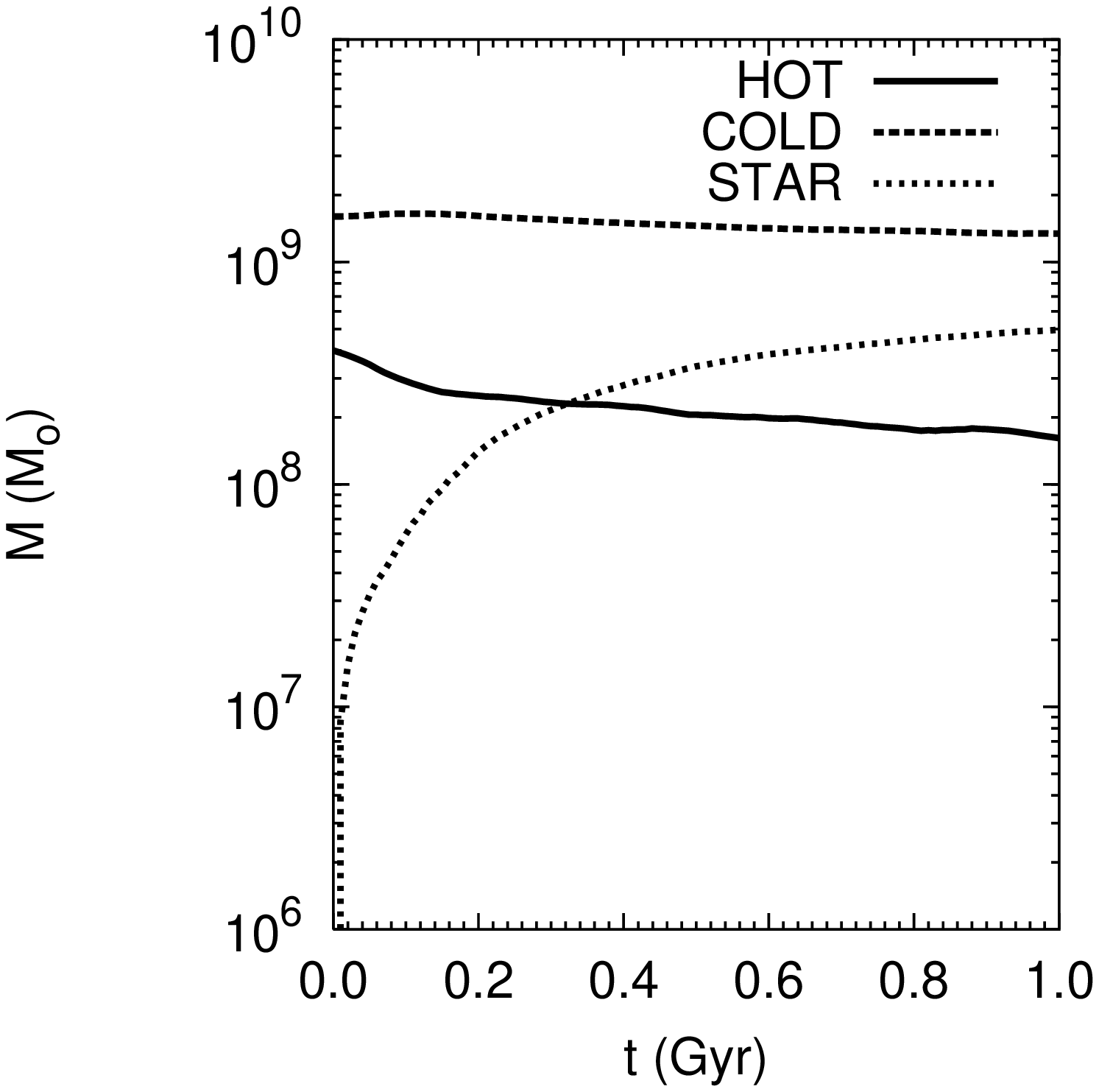} &
\includegraphics[width=2.35in]{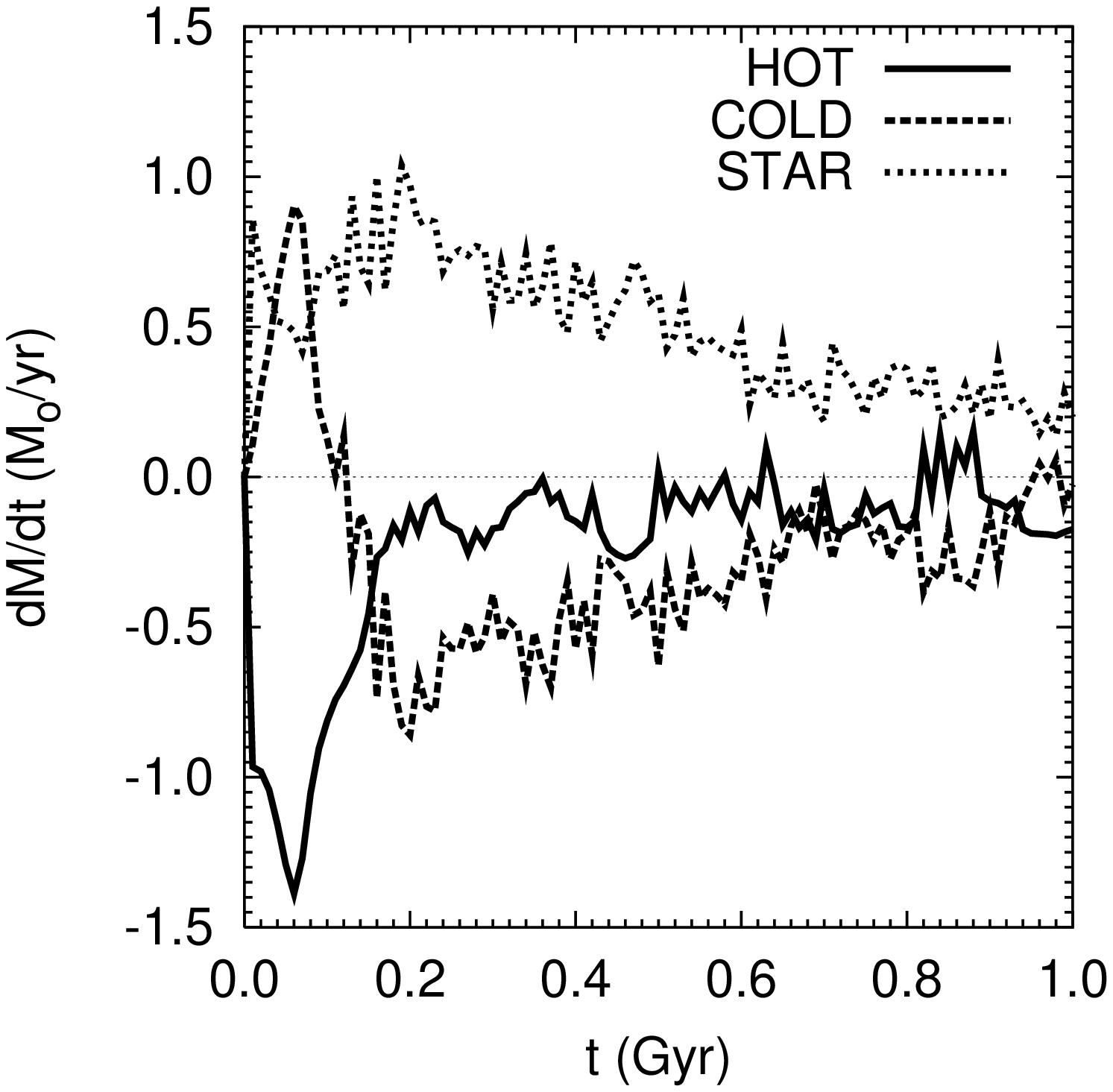}
\end{tabular}}
\caption{The temporal evolution of the mass (left) and mass exchange
rate (right) for the different components of the model galaxy.}
\label{fig:mass-t&dmdt-t}
\end{figure}

\section{Basic ingredients of the code}

For the parametric description of the cold clumps in the code we use
the mass {\it vs.} radius relation for clouds based mainly on observations
and also some theoretical work in this direction \citep{L1981,
SRBY1987, M1990}:
$$
h_{\rm cl}~\simeq~50~\cdot~\sqrt{ \frac{m_{\rm cl}}{10^6~{\rm M}_{\odot}} } {\rm \;\;\; pc}
$$

This parameterization has already successfully been applied for the
description of the cloudy medium of the ISM in \cite{TH1993, SHT1997}.

The basic mechanism for the mass exchange between ``cold'' and ``hot''
gaseous phases is a condensation {\it vs.} evaporation (CE)
of the cold cloud clumps. In our code we follow the prescription of
these processes using the model proposed in \cite{CMcKO1981, KTH1998}.
In this model the basic parameter controlling the process of CE is
$\sigma_0$, which gives the ratio between a typical length scale of
electron thermal conduction and the cloud size (compare also
\cite{McKB1990, BMcK1990}). If the cloud is small or the hot gas has a high
temperature the conduction length may exceed the cloud size
($ \sigma_0 > 1$) and conductive evaporation is limited by saturation. On
the other hand, if the temperature of the hot gas becomes very small or 
the cloud size very large, the cooling length scale becomes shorter than the
cloud size and condensation substitutes evaporation. For simplicity and
as in \cite{CMcKO1981} we just use here $\sigma_0 = 0.03$ as transition
value from evaporation to condensation, although a more detailed physical
description should invoke the cooling or field length \citep{McKB1990,
BMcK1990}. In total the rate with which ``cold'' clouds evaporate to
the surrounding ``hot'' gas their own material or acquire mass by
condensation from the surrounding gas is
$$
  \frac{dm_{\rm cl}}{dt}~=~
  \left\{
  \begin{array}{lcl}
  ~0.825~\cdot~{\rm T}_{ \rm hot}^{5/2} ~h_{\rm cl} ~\sigma_0^{-1} & \sigma_0 < 0.03 & {\rm \bf Condensation} \\
  -27.5~\cdot~{\rm T}_{ \rm hot}^{5/2} ~h_{\rm cl} ~\Phi & 0.03 \leq \sigma_0 \leq 1.0 & {\rm \bf Evaporation} \\
  -27.5~\cdot~{\rm T}_{ \rm hot}^{5/2} ~h_{\rm cl} ~\Phi ~\sigma_0^{-5/8} & \sigma_0 > 1.0 & {\rm \bf Saturated~Evap.} \\
  \end{array}
  \right.
$$

\noindent where we have used $\Phi = 1$ (no inhibition of evaporation by magnetic
fields) and
$$
\sigma_0~=~\left(~\frac{{\rm T}_{ \rm hot} ({\rm K})}{1.54 \cdot 10^7}~\right)^2
~\frac{1}{\Phi ~n_{ \rm hot} ({\rm cm}^{-3}) ~h_{\rm cl} ({\rm pc})}
$$

In our model the first important dynamical effect in the list of
interaction between the two gaseous phases is a cloud dragging (DRAG).
For this reason we use the prescription proposed in the papers
\cite{SMGYGR1972, BKS1972}.
$$
\frac{d {\bf p}_{\rm cl}}{dt}~=~C_{\rm DRAG}~\cdot~\pi h^2_{\rm cl}~\rho_{ \rm hot}~
\mid {\bf v}_{\rm cl} - {\bf v}_{ \rm hot} \mid~({\bf v}_{\rm cl} - {\bf v}_{ \rm hot})
$$

The drag coefficient $C_{\rm DRAG}$ represents the ratio of the
effective cross section of the cloud to its geometrical one $\pi
h^2_{\rm cl}$ and is set to 0.5. A value of order unity for $C_{\rm
DRAG}$ has the physically correct order of magnitude for the forces
exerted by a pressure difference before and after a supersonic shock
wave \citep{CF1998}.

The second important dynamical effect in the evolution of the cloudy
medium is a cloud {\it vs.} cloud collisions (COLL). These
processes also can significantly reduce the kinetic energy of the
cloudy system. As a first approach for these processes we assume that
in each collision the colliding clouds loss only 10 \% of its kinetic
energy.

Stars inject a lot of mass, momentum and energy (both mechanical and
thermal) in the galactic system through Super Novae (SN) explosions,
Planetary Nebula (PN) events and Stellar Wind (SW). The gas dynamics is
then strongly depend on the star formation (SF) and a feed back (FB)
processes.

Stars are supposed to be formed from collapsing and fragmenting  cold
gaseous clouds. Some possible SF criteria in numerical simulation have
been examined \citep{K1992, NW1993, FB1995}. In our code we use the
"standard" Jeans instability criterion inside the cloud particle, with
randomized efficiency for SF. As a first step we select the cloud
particles with:
$$
h_{\rm cl}~>~\lambda_{\rm J}~\equiv~c_{\rm cl}~\sqrt{ \frac{\pi}{G~\rho_{\rm cl}}}
$$

After, we calculate the maximum SFR using the idea, what the whole
Jeans mass $M_{\rm J}$ from the cloud convert to star particle during
the $\tau_{\rm cl}^{\rm ff}$ time inside the Jeans volume $V_{\rm J}$:
$$
\frac{d \rho^{\rm max}_{*}}{d t}~\equiv~
\frac{M_{\rm J}}{\tau_{\rm cl}^{\rm ff}~V_{\rm J}}~=~
\frac{4}{3}~\sqrt{\frac{6~G}{\pi}}~\cdot~\rho^{3/2}_{cl}
$$

\noindent where: $\tau_{\rm cl}^{\rm
ff}~\equiv~\sqrt{\frac{3~\pi}{32~G~\rho_{\rm cl}}}$. The actual SFR in
the each current act of SF we set by randomized these maximum SFR:
$$
\frac{d \rho_{*}}{d t}~=~{\rm \bf RAND}~(0.1~\div~1.0)~\cdot~\frac{d \rho^{\rm max}_{*}}{d t}
$$


Every new ``star'' particle in our SF scheme represents a separate,
gravitationally bound star formation macro region i.e. Single Stellar
Population (SSP). The ``star'' particle is characterized its own time
of birth $~t_{\rm SF}~$ which is set equal to the moment of particle
formation. We assume that in the moment of creation the ``star''
particle, the individual stars inside our macro ``star'' particle
distributed accordingly the \cite{KTG1993} Initial Mass Function (IMF).

\begin{figure}[t!]
\centerline{%
\begin{tabular}{c@{\hspace{0.1in}}c}
\includegraphics[width=2.35in]{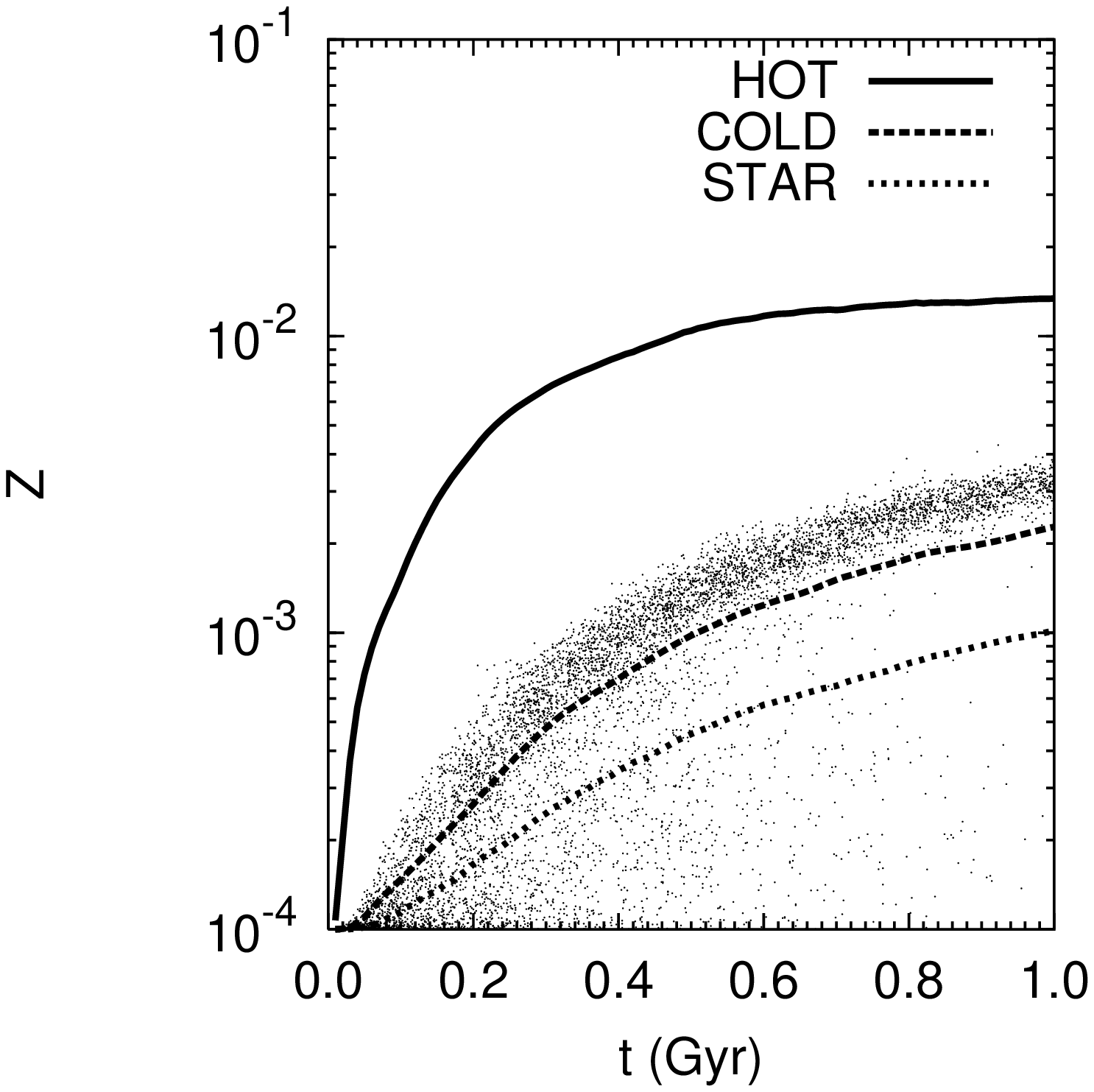} &
\includegraphics[width=2.35in]{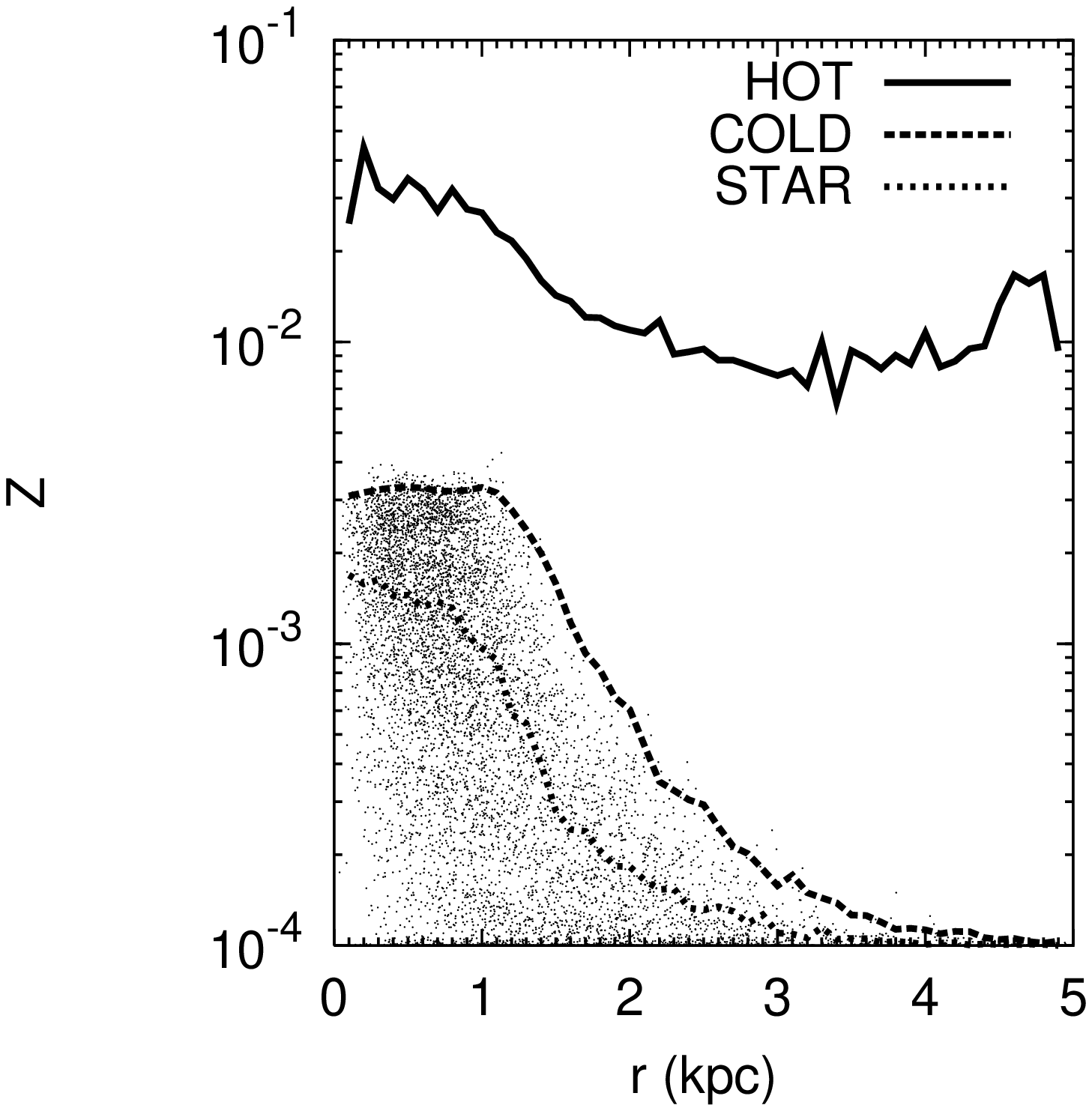}
\end{tabular}}
\caption{Temporal evolution of the metallicities (left) and their radial
distribution after 1 Gyr (right). Individual metallicities of newly born
stars are marked by dots.}
\label{fig:z-t&z-r}
\end{figure}

During the evolution, these ``star'' particles return the chemically
enriched gas to surrounding ``gas'' particles due to SNII, SNIa, PN
events. As a first approach, we consider only the production of
$^{16}$O and $^{56}$Fe. The ``star'' particles return to ISM also the
energy due to the SW, SNII, SNIa, PN processes. The total energy
released by ``star'' particles calculated at each time step and
distributed (in the form of thermal energy) between the neighbor
(N$_{\rm B}$~=~50) ``gas'' particles.

The code also includes the photometric evolution of each ``star''
particle, based on the idea of the SSP \citep{BCF1994, TCBF1996}. At
each time - step, absolute magnitudes: M$_{\rm U}$, M$_{\rm B}$,
M$_{\rm V}$, M$_{\rm R}$, M$_{\rm I}$, M$_{\rm K}$, M$_{\rm M}$ and
M$_{\rm BOL}$ are defined separately for each ``star'' particle. The
spectro - photometric evolution of the overall ensemble of ``star''
particles forms the Spectral Energy Distribution (SED) of the galaxy.

\begin{figure}[t!]
\centerline{%
\begin{tabular}{c@{\hspace{0.1in}}c}
\includegraphics[width=2.35in]{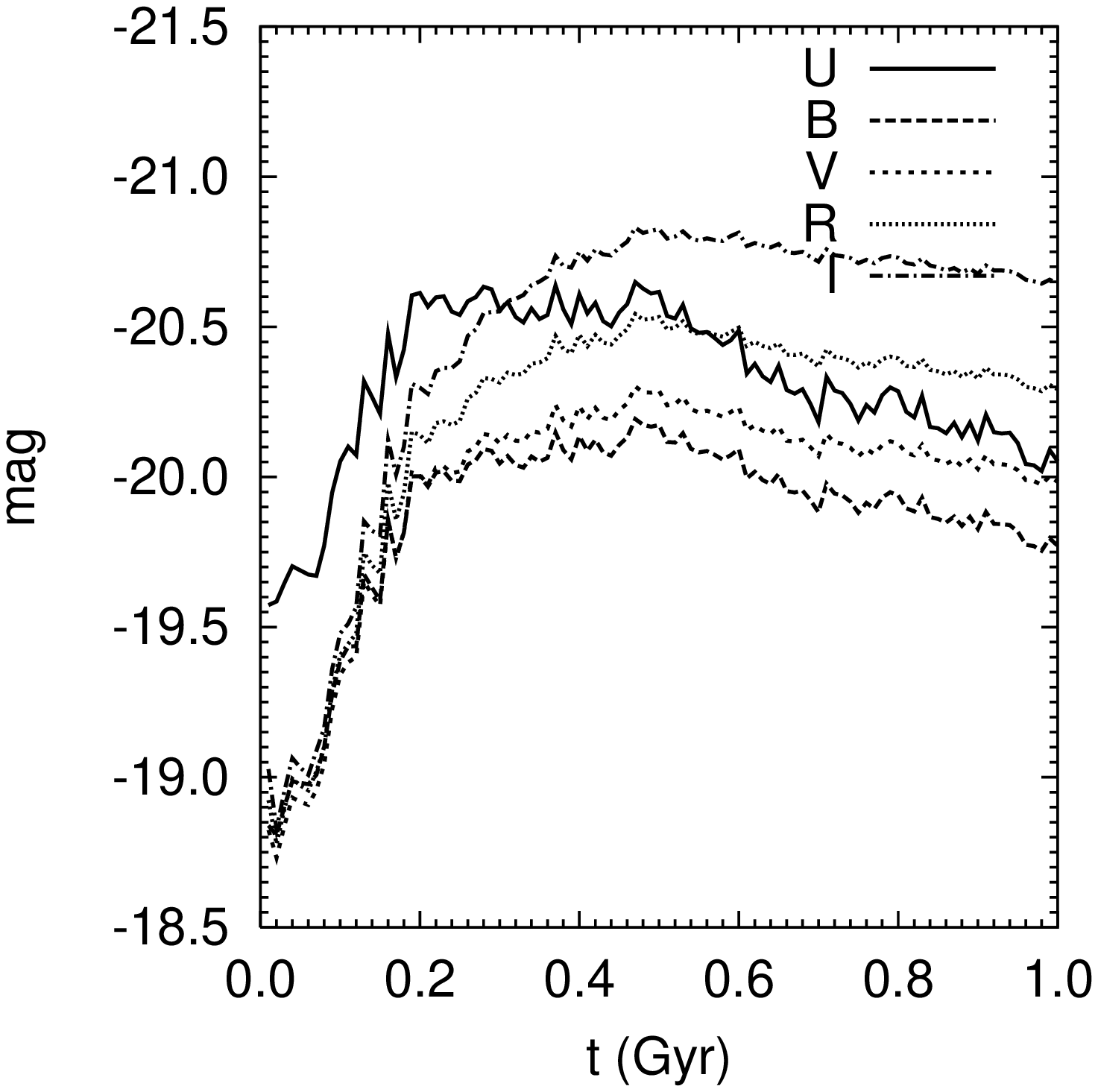} &
\includegraphics[width=2.35in]{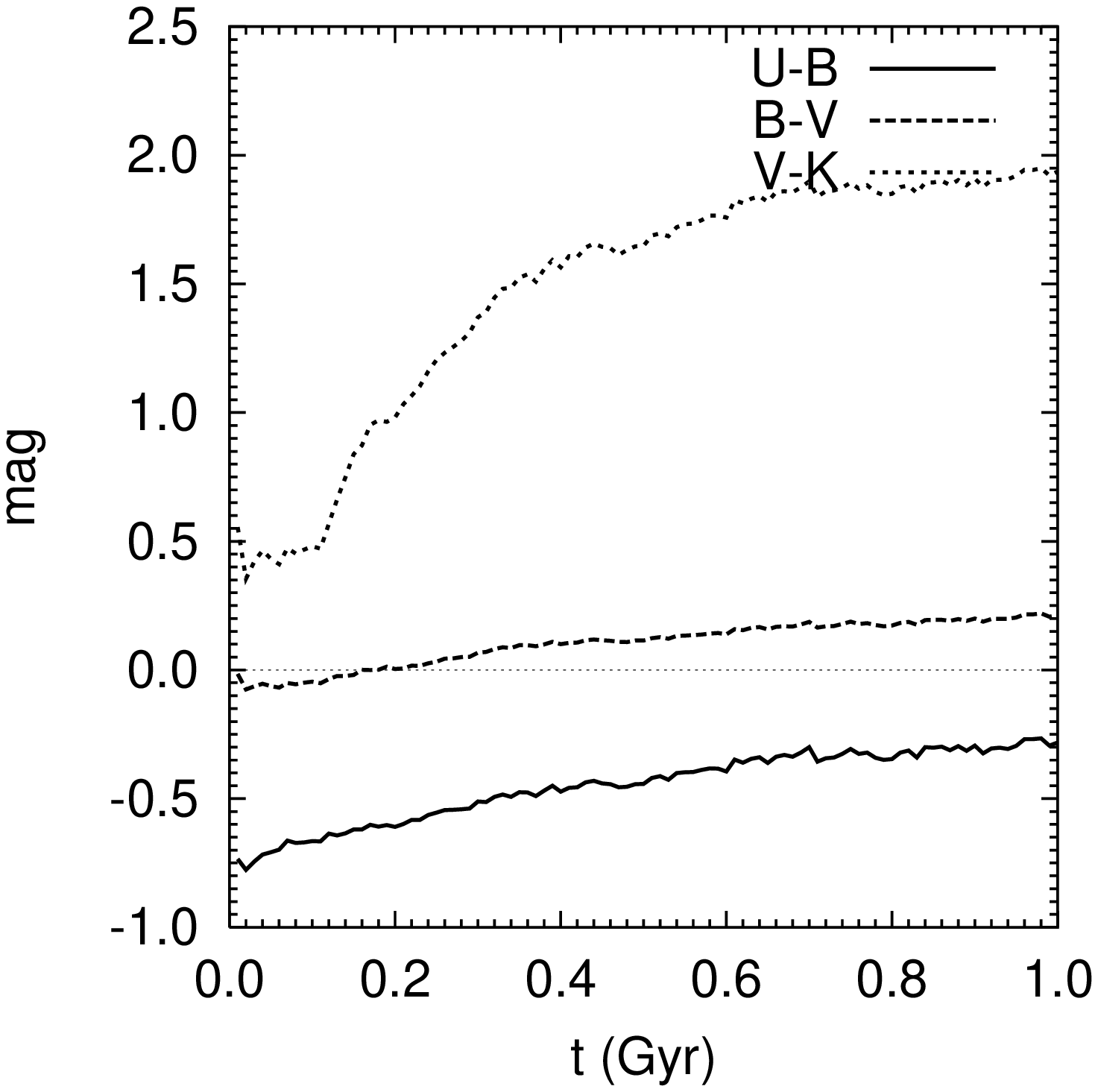}
\end{tabular}}
\caption{Temporal evolution of the model galaxy magnitudes in different
spectral branch (left) and the color indexes (right).}
\label{fig:mag-t&ci-t}
\end{figure}

\section{Initial conditions}

As a test of our new code, we calculate the evolution of an isolated
star forming dwarf galaxy. The initial total gas content of our dwarf
galaxy is $2 \times 10^9$ M$_\odot$ (80 \% ``{\tt COLD}'' + 20 \% ``{\tt
HOT}'' which is placed inside a fixed dark matter halo with parameters
r$_0$ = 2~kpc and $\rho_0 = 0.075$ M$_\odot$/pc$^3$ \citep{B1995}. With
these parameters the dark matter mass inside the initial distribution
of gas (20~kpc) is $\simeq~2 \times 10^{10}$ M$_\odot$. The initial
temperatures for the cold gas we set 10$^3$ K, for the hot gas 10$^5$
K. For the initial gas distribution we use a Plummer-Kuzmin disk with
parameters a = 0.1~kpc and b = 2~kpc \citep{MN1975}. The gas initially
rotates in centrifugal equilibrium around the z-axis.

We choose the dwarf galaxy as an appropriate object for our code,
because in this case even with a relatively ``small'' number of cold
``clouds'' ($\sim$ 10$^4$) we achieve the required physical resolution
for a realistic description of individual molecular clouds ($\sim$
10$^5$ M$_\odot$) as a separate ``{\tt COLD}'' particle. In the
simulation we use N$_{\rm hot}$ = 10$^4$ SPH and N$_{\rm cold}$ =
10$^4$ ``{\tt COLD}'' particles. After 1~Gyr more then $10^4$
additional stellar particles are created.

\section{First results}

After a moderate collapse phase the stars and the molecular clouds
follow an exponential radial distribution, whereas the diffuse gas
shows a central depression as a result of stellar feedback
Fig.~\ref{fig:100-mass&100-sigma}. The metallicities of the galactic
components behave quite differently with respect to their temporal
evolution as well as their radial distribution
Fig.~\ref{fig:mass-t&dmdt-t}. Especially, the ISM is at no stage well
mixed.

In Fig.~\ref{fig:100-mass&100-sigma}. we present the mass and surface
density distribution of the different components in the central region
of the model after 1~Gyr of evolution. In the region up to $\approx$
2~kpc the baryonic matter dominates over the DM. The surface density of
the stars can be well approximated by an exponential disk with a scale
length of 0.55~kpc. In the distribution of hot gas, we see a central
``hole'' ($\approx$ 1~kpc), as a result of gas blow-out from the center
mainly due to SN explosions but not for the cold gas. This results
disagrees with the model by \cite{MYN1999} where a density hole
occurs caused by their single gas-phase treatment.

In Fig.~\ref{fig:mass-t&dmdt-t}. we present the evolution of the mass
and the mass exchange rate of the different components. The SFR (i.e.
dM$_{\rm STAR}$/dt) peaks to a value of 1 M$_\odot$yr$^{-1}$ after 200
Myrs. Afterwards it drops down to 0.2 M$_\odot$yr$^{-1}$ within several
hundred Myrs. I.e. that after 1 Gyrs the stellar mass has already
reached $5 \times 10^8$ M$_\odot$. Another interesting feature is the
behaviour of the hot gas phase mass exchange. After the initial violent
phase of condensation an equilibrium is established which gives a hot
gas fraction of about 10\% of the total gas mass.

The metal content of the diffuse gas and the clouds differs
significantly over the whole integration time
(Fig.~\ref{fig:z-t&z-r}.). Due to SNII and SNIa events the metallicity
of the hot phase exceeds that of the clouds by almost one order of
magnitude. The clouds mainly get their metals by condensation of the
hot phase. The central metallicity plateau (up to 1~kpc) of the cold
component is explained by the fact, that condensation of metal-enriched
material does not work efficiently in that region. This signature
agrees well with the observed abundance homogeneity in dIrrs over up to
1~kpc (e.g. in I~Zw~18: \cite{I1999}). Moreover, the conditions in the
center lead mainly to evaporation of clouds which also prevents the
mixing with the metal enriched hot gas.

In the Fig.~\ref{fig:mag-t&ci-t}. we present the evolution of the model
galaxy magnitudes in different spectral branch and also the color
indexes.

\begin{acknowledgement}
The work was supported by the German Science Foundation (DFG) with the
grants 436 UKR 18/2/99, 436 UKR 17/11/99 and the SFB439 (subproject B5)
at the University of Heidelberg. P.B. is grateful for the hospitality
of the Astronomisches Rechen-Institut (Heidelberg) where the main part
of this work has been done. The calculation has been computed with the
{\tt GRAPE5} system at the Astronomical Data Analysis Center of the
National Astronomical Observatory, Japan. R.Sp. acknowledges support by
the German-Japanese cooperation grant 446 JAP 113/18/0-2.
\end{acknowledgement}

\newpage



\end{document}